\documentclass[12pt,preprint]{aastex}

\begin{document}


\title{Orbital Perturbations of the Galilean Satellites \\ During Planetary Encounters}


\author{Rogerio Deienno\altaffilmark{1} and David Nesvorn\'y}
\affil{Southwest Research Institute, Boulder, CO, United States}
\email{rogerio.deienno@gmail.com}

\author{David Vokrouhlick\'y\altaffilmark{2}}
\affil{Institute of Astronomy, Charles University, Prague, Czech Republic}

\and

\author{Tadashi Yokoyama}
\affil{Universidade Estadual Paulista, Rio Claro, SP, Brazil}


\altaffiltext{1}{{\it presently at} Instituto Nacional de Pesquisas Espaciais, S\~ao Jos\'e dos Campos, SP, Brazil}
\altaffiltext{2}{{\it visiting} Southwest Research Institute, Boulder, CO, United States}


\begin{abstract}

The Nice model of the dynamical instability and migration of the giant planets can explain many properties 
of the present Solar System, and can be used to constrain its early architecture. In the jumping-Jupiter 
version of the Nice model, required from the terrestrial planet constraint and dynamical structure of the 
asteroid belt, Jupiter has encounters with an ice giant. Here we study the survival of the Galilean 
satellites in the jumping-Jupiter model. This is an important concern because the ice-giant encounters, 
if deep enough, could dynamically perturb the orbits of the Galilean satellites, and lead to implausible 
results. We performed numerical integrations where we tracked the effect of planetary encounters on the Galilean 
moons. We considered three instability cases from \citet{NM12} that differed in the number and distribution of 
encounters. We found that in one case, where the number of close encounters was relatively small, the Galilean satellite 
orbits were not significantly affected. In the other two, the orbital eccentricities of all moons were excited 
by encounters, Callisto's semimajor axis changed, and, in a large fraction of trials, the Laplace resonance 
of the inner three moons was disrupted. The subsequent evolution by tides damps eccentricities and can recapture 
the moons in the Laplace resonance. A more important constraint is represented by the orbital inclinations 
of the moons, which can be excited during the encounters and not appreciably damped by tides. We find that 
one instability case taken from \citet{NM12} clearly fails this constraint. This shows how the regular 
satellites of Jupiter can be used to set limits on the properties of encounters in the jumping-Jupiter model, 
and help us to better understand how the early Solar System evolved.

\end{abstract}


%\keywords{}

% %%%%%%%%%%%%%%%%%%%%%%%%%%%%
\section{Introduction} \label{intro}

It is currently well accepted that the outer planets radially migrated in the past \citep{FI96}. In \citet{HM99} 
and \citet{Tsiganis05} two different migration histories were proposed. In the former, the planets were suggested to 
migrate in a smooth manner from their original orbits to those that they occupy today. The latter, instead, 
invoked a highly chaotic stage, where the outer planets underwent close encounters among themselves. Specifically, 
in the simulations of \citet{Tsiganis05}, the outer planets were initially located between 5 and 18 AU, and a massive 
outer planetesimal disk was placed beyond 20 AU. The instability was triggered in these simulations when Jupiter and 
Saturn migrated (by scattering planetesimals) over their mutual 2:1 mean motion resonance (MMR). During the 
instability, the orbits of Uranus and Neptune became Saturn-crossing, Uranus and Neptune were scattered 
out by Saturn, and these planets subsequently migrated to their current locations by gravitationally 
interacting with the outer disk. This model, also known as Nice model, appears to well explain many properties 
of the present Solar System,
such as the final orbital elements of the giant planets \citep{Tsiganis05}, 
origin of the Late Heavy Bombardment \citep{Gomes05}, 
capture of Jupiter's Trojans and the irregular satellites at Saturn, Uranus and Neptune \citep{Morbidelli05,Nesvorny07}, 
origin of the dynamical structure of the Kuiper belt, and 
the implantation of primitive trans-Neptunian objects into the outer asteroid belt \citep{Levison08,Levison09}.

However, as originally envisioned, the Nice model is unlikely to be correct in details. This is, for example, because 
the initial configuration of planets used in \citet{Tsiganis05} is difficult to reconcile with the previous stage
when planets formed and migrated in the protoplanetary gas disk. Instead, according to \citet{Morbidelli07}, Jupiter 
and Saturn should have emerged from the dispersing gas nebula with orbits in the 3:2 MMR (or possibly the 2:1 MMR), 
while Uranus and Neptune should have evolved onto nearby resonant orbits as well. The instability trigger proposed 
by \citet{Morbidelli07} was the crossing of the 5:3 MMR between Jupiter and Saturn (later the crossing of the 2:1 MMR 
also occurs, but much faster than in \citet{Tsiganis05}).
Another possible instability trigger was proposed by \citet{Levison11}, were the authors assumed 
a self-gravitating planetesimal disk. In that work, \citet{Levison11} considered the delay between the dispersion 
of the proto-solar nebula and the instability. The planets were initially assumed to be locked in a multi-resonant state \citep{Morbidelli07}, 
and the inner edge of the planetesimal disk was placed several AUs beyond the orbit of the outermost planet. 
\citet{Levison11} modeled the disk's viscous stirring, induced by the presence of Pluto-sized objects embedded in the outer disk. 
They showed that viscous stirring leads to an irreversible exchange of energy between planets and the planetesimal disk. 
This exchange of energy induces the inward migration of the inner ice giant. 
As this planet is locked in resonance with Saturn, due to the adiabatic invariance, its eccentricity increases. 
During this process, the system crosses many weak secular resonances. 
Those resonances can disrupt the mean motion resonances and make the planetary system unstable 
long time after the dispersal of the protoplanetary nebula.

Additional modifications of the original Nice model were motivated by the evolution of secular modes during planetary 
migration, mainly $g_5$, $g_6$, and $s_6$, and their effects on the terrestrial planets and asteroid belt. 
\citet{Brasser09} found that it would be problematic if $g_5$ slowly swiped over the $g_1$ or $g_2$ modes of the 
terrestrial planets, because the $g_1=g_5$ and $g_2=g_5$ resonances would produce excessive excitation and 
instabilities in the terrestrial planet system (see also Agnor \& Lin 2012). Moreover, %\citet{AL12}
\citet{Morbidelli10} and \citet{MM11} showed that the behavior of the $g_6$ and $s_6$ modes was crucially important 
for the asteroid belt, where, again, {\it slow} evolution of the $g_6$ and $s_6$ modes would violate constraints 
from the orbital distribution of asteroids. 

To avoid these problems, it has been suggested that Jupiter's orbit should have discontinuously changed by encounters 
with an ice giant. The $g_5$, $g_6$, and $s_6$ frequencies are mainly a function of the orbital separation between Jupiter 
and Saturn. Thus, as Jupiter's semimajor axis moves in during encounters (and Saturn's semimajor axis moves out),
these frequencies decrease in discrete steps. This kind of evolution is desirable, because the secular resonances with 
the terrestrial planets can be {\it stepped over}, and not activated. This instability model is known as the 
{\it jumping-Jupiter} model. Taken together, the initial conditions leading to the instability and the dynamical 
evolution of the planets were likely different from those originally envisioned in the Nice model.

\citet{Nesvorny11} and \citet{NM12} studied the possibility that more then four giant planets formed in the outer
Solar System. They showed that including a planet with mass comparable to that of Uranus or Neptune on an orbit
between the original orbits of Saturn and Uranus can significantly increase the success rate of instability
simulations. This is because, more often than not at least one ice giant is ejected from the Solar System during 
the instability. The five-planet cases considered in \citet{Nesvorny11} and \citet{NM12} showed just the right kind of 
the jumping-Jupiter evolution discussed above, and also often satisfied various other constraints. The six-planet case 
worked as well but did not offer significant advantages over the five-planet case.

To understand the early evolution of the Solar System it is important to not only consider the evolution of giant
planets, but also to determine the effects of the planetary evolution on the populations of small bodies.
Many recent works studied the effects of planetary migration on Jupiter and Neptune Trojans, asteroids and Kuiper 
belt objects, etc. \citep{Morbidelli05, Levison08, Levison09, NV09}. Here we consider the planetary satellites.

The satellites of the giant planets can be divided into several categories. The {\it regular} moons have orbits close 
to their host planet, and small orbital eccentricities and orbital inclinations with respect to the planet's equator.
The {\it irregular} satellites, on the other hand, have distant orbits, and high eccentricities and high inclinations 
(as measured with respect to planet's orbital plane). It is believed that the irregular satellites were captured from 
heliocentric orbits. \citet{Nesvorny07}, for example, suggested that the irregular satellites were captured during encounters
between planets in the Nice model when background planetesimals were deflected onto bound orbits. They showed that 
this type of capture has the right efficiency (up to a factor of few) to explain observations and leads to a roughly 
correct distribution of orbits of captured satellites at Saturn, Uranus and Neptune. However, because Jupiter does not 
generally participate in planetary encounters in the original Nice model, \citep{Nesvorny07}, where the encounter 
statistics was based on the original Nice model simulations, were unable to address with their model the origin of the 
irregular satellites at Jupiter. The problem of the irregular satellite capture at Jupiter was recently reconsidered 
by us \citep{NVD14} in the context of the jumping-Jupiter model, where Jupiter participates in encounters. 

The regular satellites represent a different concern. They presumably formed near their present orbits well before
the instability, and were exposed to the effects of planetary encounters during the instability. Therefore, it is 
interesting to evaluate the degree of orbital excitation to determine whether the currently-favored instability 
models are consistent with the systems of regular satellites we see at the outer planets today. 

Our work described here builds on the previous efforts that considered the effect of planetary encounters on the regular 
satellites. \citet{Deienno11} studied the history of planetary encounters in the instability model with four planets 
and found that the regular satellites at Uranus became destabilized in about 40\% of the considered cases, if 
Uranus had close encounters with Saturn. Also, if these encounters occurred, any satellites beyond Oberon's orbit 
would have most certainly become unbound \citep{Deienno11}.

The encounters between Uranus and Saturn do not typically occur in the instability model with five planets (Nesvorn\'y 
\& Morbidelli 2012). The survival of Uranus's satellites is therefore less of a problem in the five-planet model   
(Nesvorn\'y et al. 2014b). Instead, in this model, Saturn has encounters with the ejected ice giant. 
Interestingly, these planetary encounters can lead to the orbital excitation of Saturn's moons Titan and Iapetus, and may
potentially explain, as shown in \citet{NVD14b}, the anomalous inclination of Iapetus with respect to the Laplace 
surface \citep{tremaine09}.

Our main goal in this paper is to determine the orbital perturbations of the Galilean moons at Jupiter in the five-planet 
jumping-Jupiter models taken from \citet{NM12}. This is an important issue because the third ice giant included
in these simulations has many dozens of close encounters with Jupiter, and is ejected as a results of these encounters 
to interstellar space. The other planets that remain in the Solar System (Saturn, Uranus and Neptune) suffer fewer
encounters with the ice giant; the perturbations of their regular satellites should therefore be less of an issue
(see Nesvorn\'y et al. 2014b). %\citet{NVD14b}
We also simulate the tidal evolution of the Galilean moons after the instability 
to show that their orbital eccentricities, if excited by the encounters, could have decreased by tides during the subsequent 
evolution.  

The structure of this paper is as follows. Section 2 explains our method for tracking the planetary encounters and 
orbits of the regular satellites. In section 3, we discuss the number and distribution of encounters in different
instability cases and the orbital excitation of the Galilean moons during encounters. We also discuss in this section
the tidal effects on the Galilean satellites and the behavior of the Laplace resonance. In Section 4, we summarize 
the main conclusions of this paper.

\section{Methodology} \label{method}
% %%%%%%%%%%%%%%%%%%%%%%%%%%%%

We used three different cases of planetary instability taken from the simulations of \citet{NM12}.  
We refer to these cases as $Case~1$, $Case~2$, and $Case~3$. They are the same cases as the ones 
used in the previous studies of capture of Jupiter's Trojans and irregular satellites (Nesvorn\'y et al. 2013, 2014a). %\citep{NVM13,NVD14}. 
By using the same cases as in the previous works we aim at subjecting the selected instability scenarios
to a number of tests. Our hope is to advance toward a fully self-consistent model of the early Solar System 
instability.

The three selected cases feature three different histories of encounters of Jupiter with ice giants, resulting in 
different dynamical perturbations of the Galilean satellites. We have considered planetary encounters whenever the 
distance between the planets was less than the sum of their Hill radii  ($d < R_{Hill}^{Jup} + R_{Hill}^{Ice}$). 
The original simulations in \citet{NM12} were repeated and we recorded all encounters of Jupiter that satisfied
the above criteria. See section \ref{enc_geometry} for a discussion of encounters in each case.

The configuration of the Galilean satellites at the onset of the planetary instability is unknown. To set up our 
simulations, we opted for using the current orbits of Io, Europa, and Ganymede from JPL 
Horizons\footnote{telnet://horizons.jpl.nasa.gov:6775 (terminal access)} 
at ten slightly different epochs. This choice implies that our initial satellite configurations
have the inner three moons in the Laplace resonance, and we can test, among other things, how the Laplace 
resonance is affected by planetary encounters. Callisto was placed on an orbit with Callisto's present semimajor 
axis, and zero eccentricity and zero inclination to Jupiter's equator. To increase the statistics, we considered 
a hundred different positions of Callisto along the orbit for each one of the ten epochs.
The mean longitude $\ell$ was set such that
$\ell=k\Delta \ell$, where integer $k=0,1,...,99$ and $\Delta \ell=3.6\degr$.    
Thus, in total, we have 1000 different configurations of the Galilean satellites for each case. 
In addition, we also considered an additional set of 1000 initial configurations, where Callisto was placed
in the outer 2:1 MMR with Ganymede. This was done to test whether it is possible that all four Galilean
satellites started in a chain of mean motion resonances, and Callisto was kicked out of the resonance with
Ganymede by planetary encounters. 

We proceed as follows:
($i$) Starting from the planetary positions and velocities recorded at the first encounter, the system containing 
the Sun and planets is integrated backward until the separation of Jupiter and the ice giant reaches 2 AU 
(let $\Delta t$ be the time for this to happen). 
($ii$) The Galilean satellites are placed at Jupiter on orbits described above.
($iii$) We then integrate the orbits of planets and the Galilean satellites forward for time $2\Delta t + P_J$, 
where $P_J$ is the orbital period of Jupiter 
(we also tested timespan of 10 and 100 times $P_J$ to verify that the results do not depend on our assumptions). 
The effects of Jupiter's oblateness ($J_2$) and obliquity ($\varepsilon\simeq3\degr$)  
are included in these integrations (precession of Jupiter's spin vector is ignored).
($iv$) The orbits of the Galilean satellites obtained at this point are used as the initial orbits for the 
next encounter and the procedure is iterated over all encounters.
($v$) Finally, the satellite orbits after the last encounter are integrated for additional 1000 years. 
We use this simulation to compute the mean orbital elements of the Galilean satellites. 
As the encounters happen in a narrow window of time, tidal damping in between encounters is negligible.
All the integrations described 
above were conducted using the Bulirsch-Stoer integrator from the Mercury code \citep{Chambers99}.

Before the computed mean elements can be compared with the real mean elements of the Galilean satellites 
in the present Solar System we must account for the orbital evolution of the Galilean satellites
from the time of instability to the present. Considering that the planetary instability most likely occurred about 
4 Gyr ago, the {\it tidal} effects can be particularly important. To study tidal evolution we modified
the symplectic integrator known as 
{\tt swift rmvs3}  \citep{Levison94} to include the tidal acceleration terms 
from \citet{Mignard79}. The tidal dissipation in this model depends the Love number $k_2$ and dissipation 
parameter $Q$. The values of these parameters for Jupiter and the Galilean satellites are poorly known. 
\citet{Lainey09} found, by fitting a tidal model to the astrometric observations of the Galilean satellites, 
that $k_2/Q=1.102 \times 10^{-5}$ for Jupiter and $k_2/Q = 0.015$ for Io. We adopt these values here, 
and for simplicity keep them constant throughout the evolution. 
The results of our tidal simulations are presented in section \ref{tides}.

 %%%%%%%%%%%%%%%%%%%%%%%%%%%%
\section{Results} \label{results}

\subsection{Properties of encounters} \label{enc_geometry}

\begin{figure}
\epsscale{1.0}
\plotone{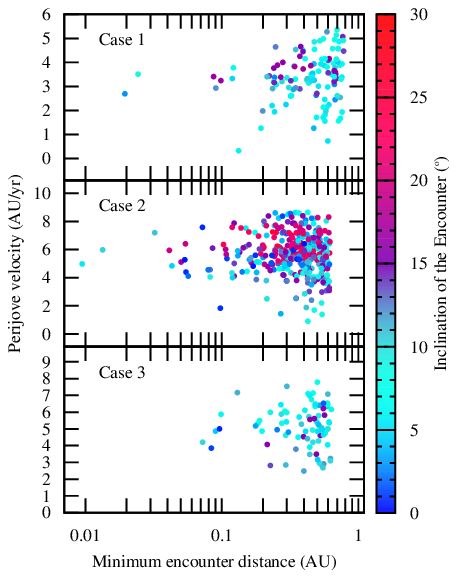}
\caption{Perijove velocity, inclination (with respect to Jupiter's orbit) and minimum distance of each recorded encounter in 
$Cases$ 1, 2 and 3. \label{fig1}}
\end{figure}

Figure \ref{fig1} shows the minimum distance between planets, perijove velocity of the ice giant,
and trajectory inclination for all planetary encounters of Jupiter. The number and geometry of 
encounters differ from case to case. The number of encounters is 117, 386 and 80 in $Cases$ 1, 2 
and 3, respectively. The number of encounters reaching minimum distance $d_{min}<0.05$ AU is 2, 9 and 0 
in these cases. As we will see below these deep encounters matter the most for the excitation of
the Galilean satellites. Most encounters have perijove velocities roughly within the range from 
3 to 6 km s$^{-1}$ and inclinations smaller than 10$^\circ$. Only $Case~2$ shows inclinations up 
to 30$\degr$. 

\begin{figure}
\epsscale{1.0}
\plotone{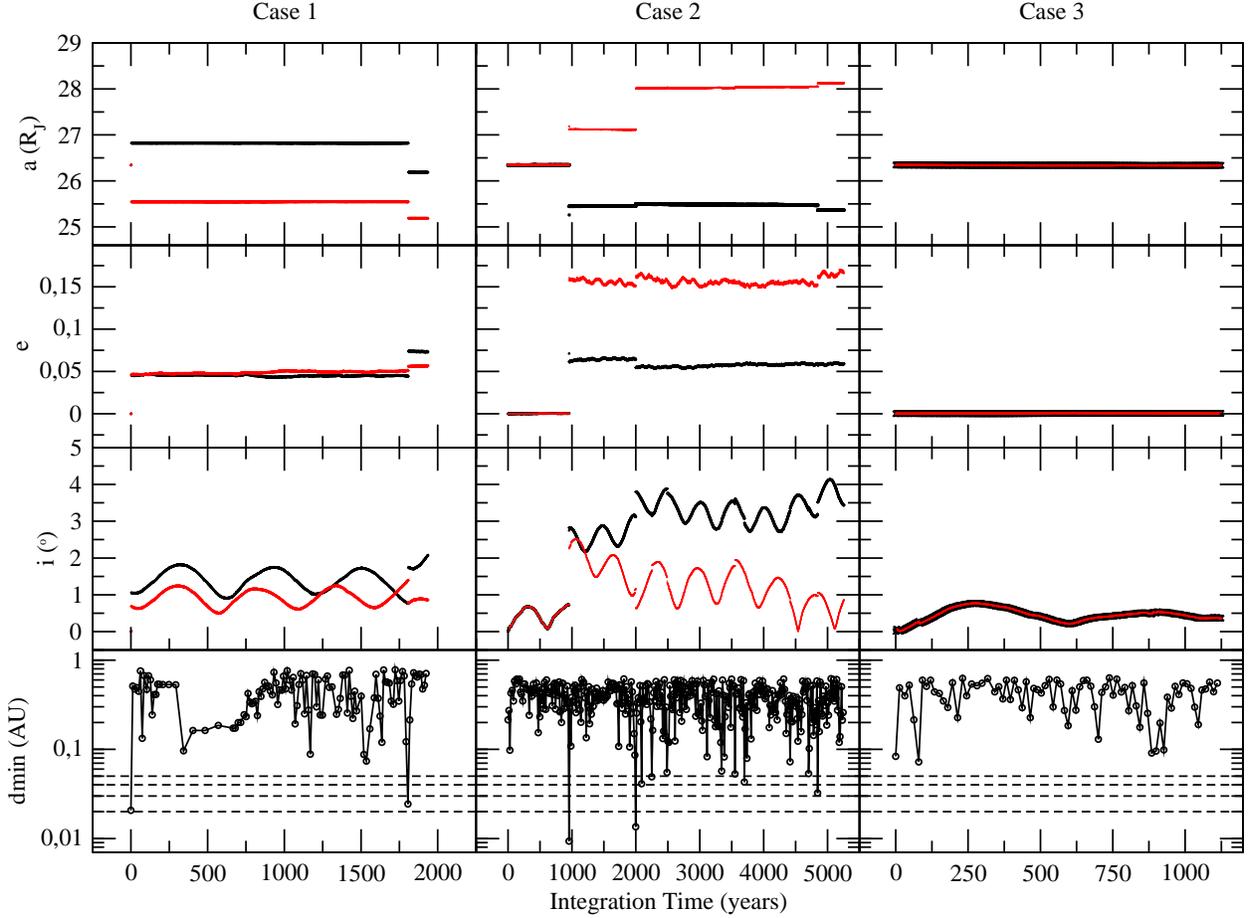}
\caption{From top to bottom: Evolution of the semimajor axis, eccentricity, orbital inclination 
(relative to the equatorial plane of Jupiter), and the minimum distance of encounters as a function of 
the integration time (integration Time = $n\times[2\Delta t+P_J], n=1...N_{enc}$), 
for two different phases of Callisto on its orbit (red and black). 
From left to right: $Cases$ 1, 2, and 3. The dashed lines in the bottom panels, from top to bottom, 
represent four values of minimum encounter distance from 0.05 AU to 0.02 AU.\label{fig2}}
\end{figure}

Figure \ref{fig2} shows an example of orbital perturbations of Callisto's orbit obtained in our simulations.
In each case, the results are shown for two different initial phases of Callisto. The figure illustrates 
that the effect of encounters can lead to different outcomes. In some encounters, the satellite eccentricity, 
inclination, and semimajor axis increase, while in other encounters the values of orbital elements decrease. 

The degree of perturbation during an encounter apparently depends on the minimal distance $d_{min}$.
In general, the encounters with $d_{min}>0.05$ AU do not cause significant variations (see $Case~3$ 
in Fig. \ref{fig2}). The encounters with $0.03 < d_{min} < 0.05$ AU can lead to relatively small changes 
in the orbital inclinations, but otherwise do not affect orbital elements much. The encounters with 
$0.02 < d_{min} < 0.03$ AU, on the other hand, are already deep enough to significantly perturb
the orbits of satellites, mainly that of Callisto (for reference, Callisto's semimajor axis is 
$\simeq$0.012 AU). Only a very few of these encounters can happen if major excitation of Callisto's
orbit is to be avoided. Finally, the encounters with $d_{min} < 0.02$ AU are the most destructive 
in that they lead to a major excitation of the satellite orbits, and, in some cases, can eject Callisto
from the system (see below).

\subsection{Excitation of satellite orbits} \label{cases}

\subsubsection{Case 1} \label{case1}

\begin{figure}
\epsscale{1.0}
\plotone{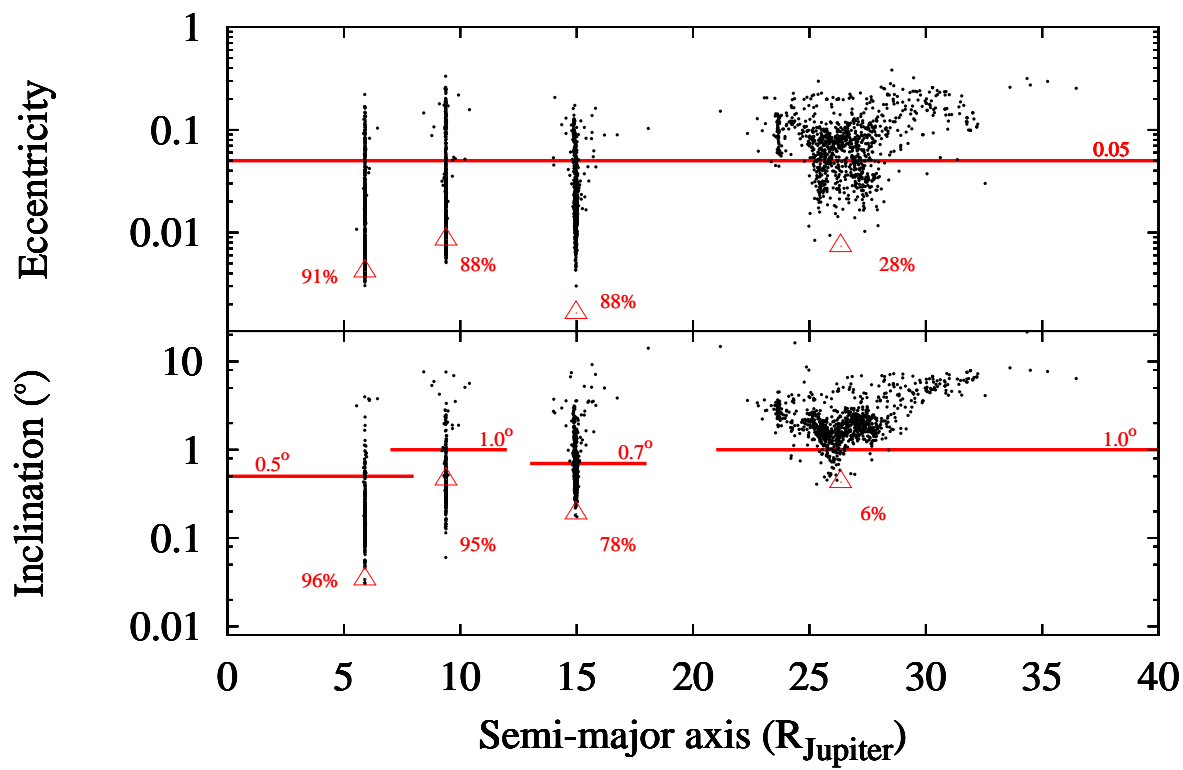}
\caption{$Case~1$. Top: Final averaged eccentricity as function of final averaged semi-major axis. 
Bottom: The final averaged orbital inclination with respect to Jupiter's equator. The simulation 
results are shown by black dots. The triangles show 
the current averaged values of the orbital elements of the Galilean moons (from left to right: Io, 
Europa, Ganymede and Callisto). The labels denote the percentage of simulated trials that ended up 
below the reference lines (discussed in the main text). \label{fig3}}
\end{figure}

The $Case~1$ is an intermediate case with 117 total encounters, two of which have $0.02< d_{min} < 0.03$ AU
(Figures \ref{fig1} and \ref{fig2}). Figure \ref{fig3} summarizes the results of our simulations in 
$Case~1$. For reference, as a simple criterion of the plausibility of the results, we plot in this figure 
the lines $e_{ref}=0.05$ for eccentricities, and $i_{ref}$ for inclinations, where $i_{ref}$ was chosen to be 
$0.5\degr$ greater than the current mean inclination. We calculate the fraction of final orbits that end up 
having $e<e_{ref}$ and $i<i_{ref}$. These fractions are shown in Figure \ref{fig3}. 

The orbits of the inner three satellites are only modestly excited by the encounters. The fraction of trials
in which the eccentricities and inclinations end up below the reference values is large ($>75$\% and 
$\gtrsim$ 90\%, respectively). These results are plausible. A more stringent constraint is represented, as expected, 
by Callisto's orbit. We find that about one third of the trials end up with Callisto's eccentricity $e<e_{ref}$. This is, 
however, still a relatively large fraction. Moreover, as we will see in section \ref{tides}, Callisto's and other 
moon's eccentricities can be damped by tides after the stage of planetary encounters. The eccentricity excitation
cannot therefore be used, in general, to rule out specific encounter histories (except if orbits become unbound). 
The results for Callisto's inclination are more interesting, because only a very small fraction of trials end up with 
$i<i_{ref}$ (Figure \ref{fig3}), and the inclination remains nearly unchanged during the subsequent tidal 
evolution (section \ref{tides}).
 
As for the effect of encounters on the semimajor axis, the inner three satellites end up very near their original 
semimajor axis values. Even if the changes are small they can have a major consequence for the Laplace resonance in that, 
at least in some cases, the libration amplitude of the Laplace resonance can significantly increase or the moons can 
end up on non-resonant orbits. We will discuss this issue in section \ref{laplace}. 

The semimajor axis of Callisto can change substantially (by up to $\simeq$5 $R_{Jupiter}$, where Jupiter's equatorial 
radius $R_{Jupiter} \simeq 71500$ km). This raises a question of whether Callisto could have originally shared 
the Laplace resonance with the inner moons (the 2:1 MMR with Ganymede is located at $a \simeq 23.76$ $R_{Jupiter}$), 
and was scattered to its current orbit ($a \simeq 26.33$ $R_{Jupiter}$) by encounters. We address this issue in 
section \ref{appendixA}.

\subsubsection{Case 2} \label{case2}

\begin{figure}
\epsscale{1.0}
\plotone{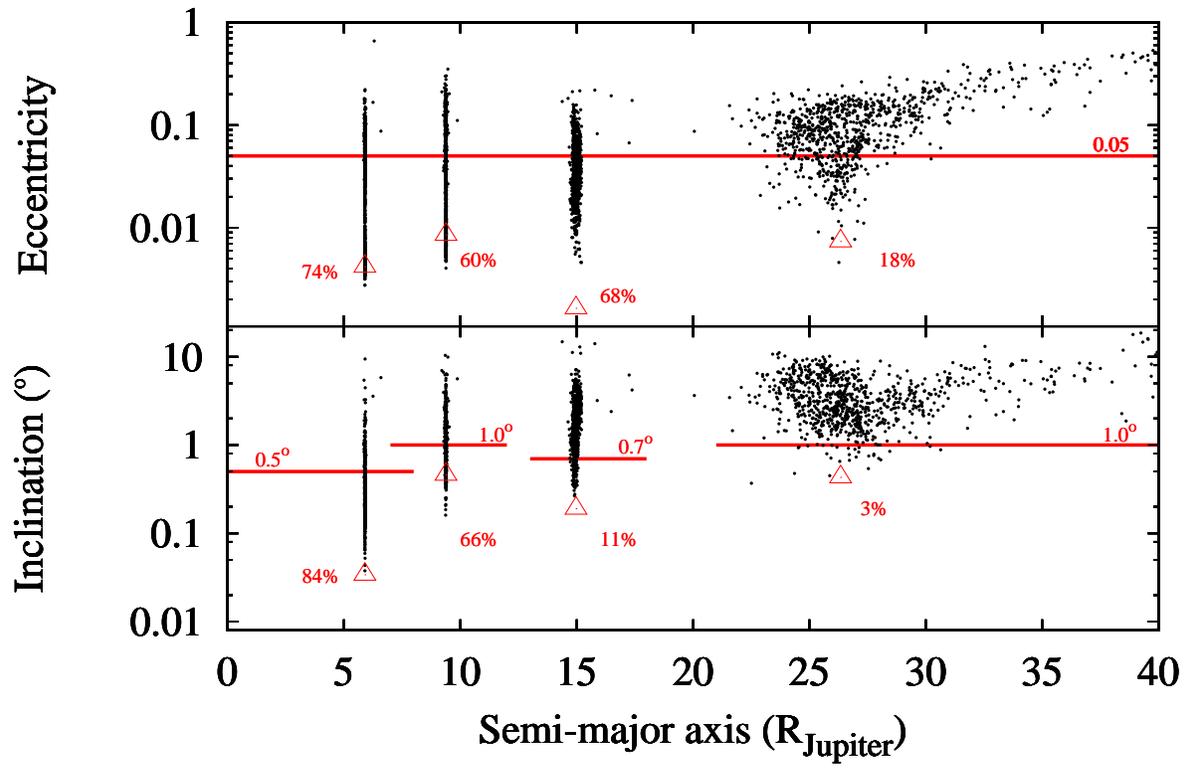}
\caption{The same as Figure \ref{fig3} but for $Case~2$. \label{fig4}}
\end{figure}

Figure \ref{fig4} shows orbital perturbations of the Galilean satellites in $Case~2$. In this case, we have  
nine encounters with $d_{min}<0.05$ AU and two encounters with $d_{min}<0.02$ AU. These close encounters produce
very large perturbations of orbits of the Galilean satellites. For example, in only 11\% of trials Ganymede's 
orbital inclination ends up below our reference value ($i<0.7\degr$), and in only 3\% of trials this criterion 
holds for Callisto's inclination. The changes on Io's and Europa's orbits are also substantial. Given these results,
we believe that this case can be ruled out. 

\subsubsection{Case 3} \label{case3}

\begin{figure}
\epsscale{1.0}
\plotone{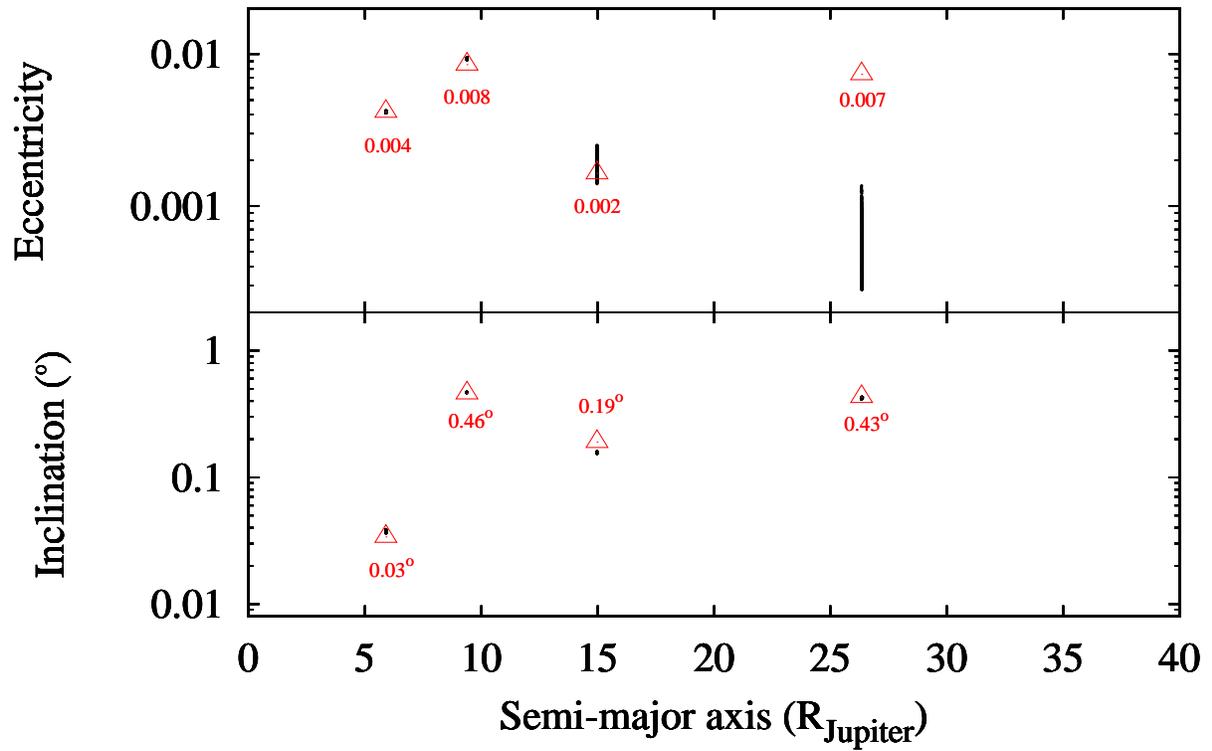}
\caption{The same as Figure \ref{fig3} but for $Case~3$. The numbers denote the average 
values of the eccentricity and inclination (with respect to the planet equator) of the Galilean moons. \label{fig5}}
\end{figure}

Figure \ref{fig5} summarizes the outcomes of our simulations in $Case~3$. In this case, there were no planetary 
encounters with $d_{min}<0.05$ AU and orbits of the Galilean moons did not change much. Therefore, $Case~3$ is
clearly plausible and would imply that the architecture of the Galilean system has remained nearly unchanged
during the instability.

An interesting observation, unrelated to planetary encounters, concerns Callisto's inclination. Recall that we 
started with Callisto's orbit with zero inclination with respect to Jupiter's equatorial plane. It may then seem 
surprising, as seen in Figure \ref{fig5}, that mean orbital inclination with respect to the equatorial plane 
ends up exactly matching the present value of Callisto ($i=0.43\degr$). This result arises because the Laplace 
surface at the location of Callisto's orbit is inclined by $\simeq 0.43\degr$ to Jupiter's equator
\citep{wardcanup06,tremaine09}. 
Thus, an orbit starting on the equatorial plane will have oscillations, 
with amplitude reaching almost $1\degr$ in the equatorial reference plane, and an average inclination value 
equal to the tilt between equatorial and Laplace planes. 
On the other hand, the oscillation of Callisto's orbital inclination with respect to the Laplace plane is negligible and follows the evolution  
of the Laplace surface during the evolution of Jupiter's semimajor axis.
This observation may have interesting implications for the origin of Callisto's
orbital inclination, because it shows that Callisto's orbit should have initially coincided with Jupiter's 
equator, as expected, for example, in the presence of a heavy circumplanetary disk, or if Jupiter's obliquity
was zero. This would require that the circumplanetary disk was dispersed, or the Jupiter's obliquity
was tilted, on a very short timescale ($\sim10^2$ yr). 

\subsection{Tests with Callisto in a resonance} \label{appendixA}

With Callisto's semimajor axis considerably changing in $Cases~1$ and $2$, we decided to test a possibility
that Callisto's orbit was initially in the 2:1 MMR with Ganymede (potentially driven there when 
satellites radially migrated by interacting with the circumplanetary gas nebula, or when 
Ganymede was pushed out by tides \citep{Yoder81,Canup02,Canup06,Canup09}), and Callisto was kicked out of 
the resonance during planetary encounters. These tests were done for $Case~1$, because this case leads 
to a large variation of Callisto's semimajor axis, as needed for moving Callisto's orbit from the resonance 
to its current location 
(without any large eccentricity and/or inclination excitation seen in $Case~2$).

\begin{figure}
\epsscale{1.0}
\plotone{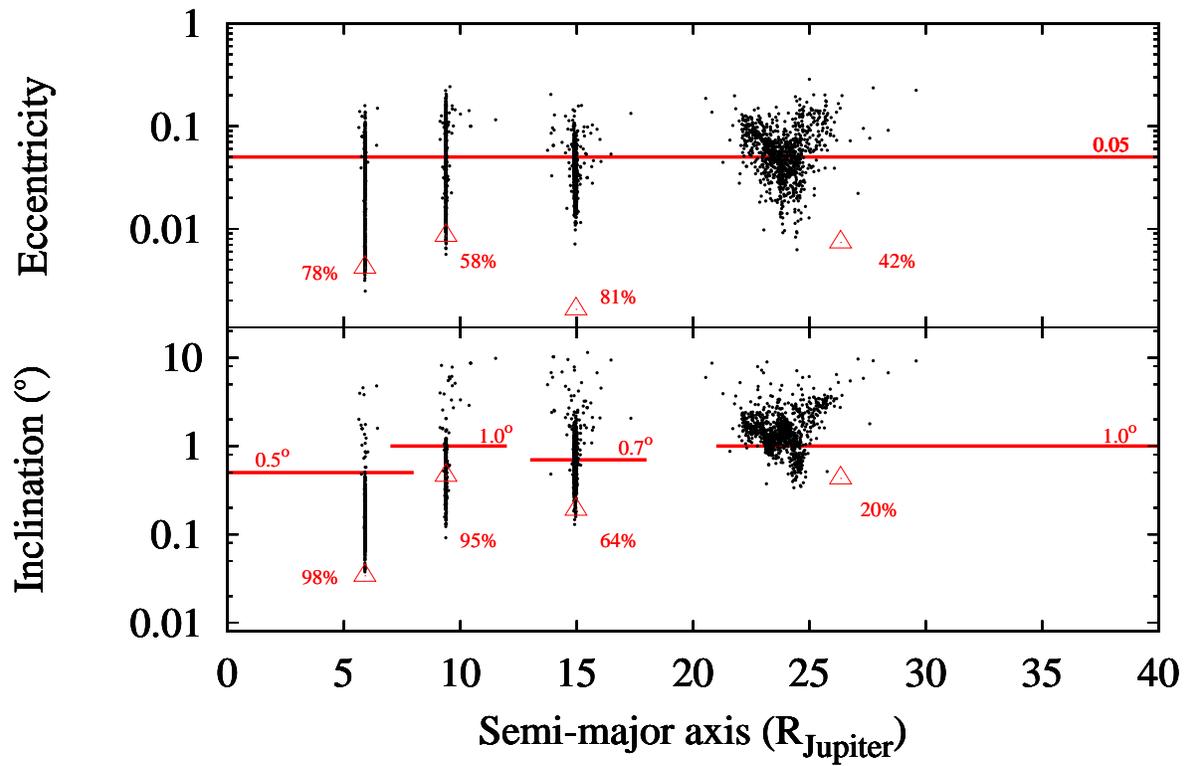}
\caption{Same as Figure \ref{fig3} but starting with all four Galilean satellites in a chain of
the 2:1 MMRs (i.e., Callisto in the 2:1 MMR with Ganymede). \label{fig9}}
\end{figure}

The initial orbits of the Galilean satellites were set by the method described in section 
\ref{method}, except that Callisto was placed in the 2:1 MMR orbit with Ganymede.   
We find that these new simulations produce results very similar to those obtained with the original
initial conditions (compare new Figure \ref{fig9} with Figure \ref{fig3}). The fraction of trials
in which Callisto's eccentricity and inclination ended up below the reference values are now 
higher than before (42\% and 20\%, respectively). This is related to a strong gradient 
of the excitation pattern with the radial distance from Jupiter and the fact that Callisto started
with a slightly smaller semimajor axis value in these new simulations.

Figure \ref{fig9} shows that, while it is plausible to change Callisto's semimajor axis by the 
required amount, the eccentricity and inclination are typically excited quite a bit 
for encounter distances $0.02 < d_{min} < 0.03$ AU, 
such that 
only a negligible fraction of trials end up with $a\simeq27$ $R_{\rm Jupiter}$, and $e<e_{ref}$
and $i<i_{ref}$. This may not be such a problem as far as the eccentricity is concerned, because
Callisto's eccentricity could have been damped by tides during the subsequent evolution (see Section \ref{tides}). 

Its orbital inclination, however, is almost unaffected by tides, and thus presents a more rigid
constraint. In particular, we do not see how the excited orbits with $a\simeq27$ $R_{\rm Jupiter}$ and
$i>3\degr$ (Figure \ref{fig9}) could reach $i<1\degr$ to become a plausible proxy for the current
Callisto's orbit. We conclude that it is unlikely Callisto could have reached its current orbit 
by starting in the 2:1 MMR with Ganymede, being kicked out of the resonance by planetary encounters,
and subsequently evolving by tides. Additional processes responsible for damping Callisto's 
inclination would have to be identified for this possibility to become viable.

\subsection{Tidal evolution} \label{tides}

In the previous sections, we determined orbital perturbations of the Galilean satellites during close 
encounters in the $jumping$-Jupiter model. These perturbations happened most likely some 4 Gyr ago.
A question therefore arises of whether (and how) the satellite orbits could have changed during the 
$\sim$4 Gyr period between the instability and the present time. Here we consider the orbital changes 
produced by the tidal interaction of moons with Jupiter. 

To this goal, we performed a number of numerical integrations with a symplectic $N$-body code known as 
{\tt swift rmvs3}  \citep{Levison94} that we modified to include the tidal acceleration terms from
\citet{Mignard79} (see also \citet{Lainey09}, equations 1 and 2 in their supplement). These simulations 
were done in a reference system centered on Jupiter. We considered both the planetary and satellites 
tides. We found that the tidal dissipation in the system is overwhelmingly due to the interaction
between Io and Jupiter. The direct tidal effects of other satellites can be neglected, but important
indirect effects arise on these satellites because they are coupled to the Jupiter-Io pair. 
Therefore, Callisto's orbital eccentricity can be damped as a consequence of the strong dissipation in the pair Jupiter-Io, 
similar to what happens in the case of extra-solar planets, when damping is applied to the innermost 
planet in a planetary system \citep{Lovis11,Van13}.

The satellite rotation was assumed to be synchronous. To implement the synchronous rotation in the code 
we adopted the following approximation (V. Lainey, personal communication). First, we considered only 
the radial component of the tidal acceleration that results from changing planetocentric distance 
(and the related dissipation). The radial component does not depend on satellite's rotation rate, and is 
therefore independent of the detailed assumptions about synchronicity. 
The longitudinal tidal acceleration
is then effectively included by multiplying the strength of the radial tide by 7/3. This is because in 
the limit of small eccentricities, which is applicable here, 
the orbital energy dissipated in satellite's librations is
4/3 of that dissipated in radial flexing (e.g., Murray \& Dermott, 1999, %\citet{Murray99}, 
Chapter 4).

The strength of tidal dissipation is parametrized by the time delay due to tidal response of the deformed 
body or equivalently by $k_2/Q$, where $Q$ is the dissipation function and $k_2$ is the Love number. 
We assume $k_2/Q = 1.102 \times 10^{-5}$ for Jupiter and $k_2/Q = 0.015$ for Io \citep{Lainey09}, 
and evolve the system for 4 Gyr. Possible variation of $k_2/Q$ over the age of the Solar System are 
not considered, because we are not interested in detailed modeling the tidal evolution, but rather 
in the average magnitude of the tidal effect in 4 Gyr.

\begin{figure}
\epsscale{1.0}
\plotone{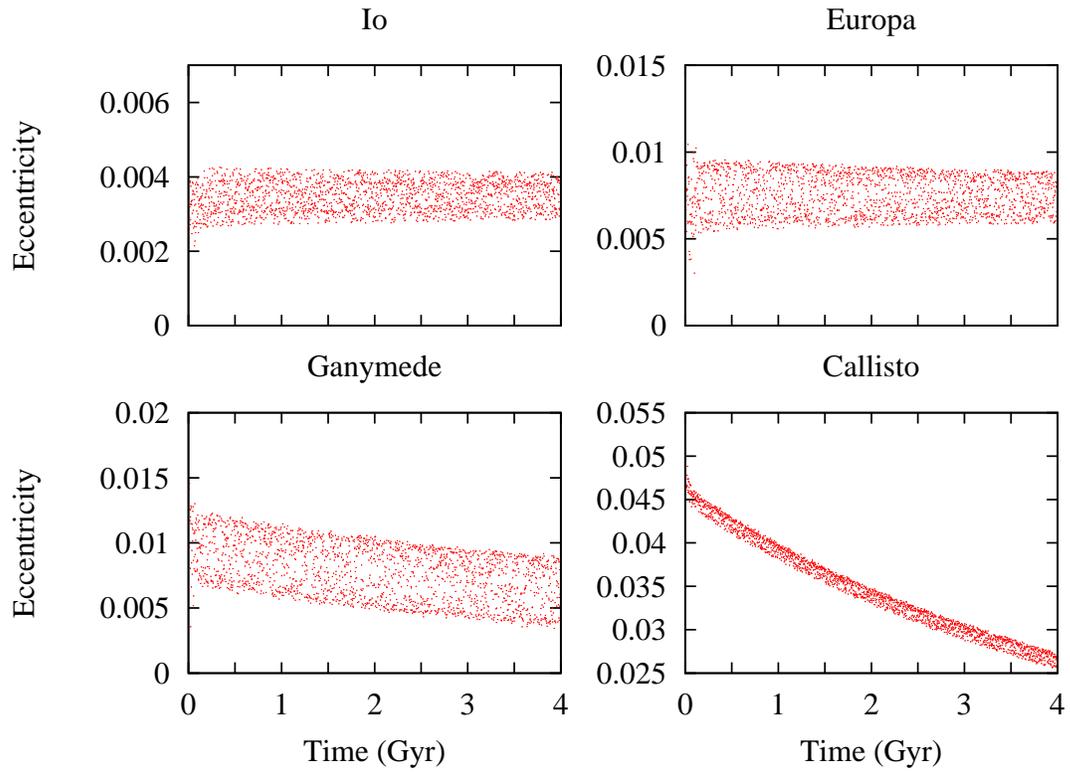}
\caption{Evolution of the orbital eccentricity of the Galilean satellites during the tidal evolution.\label{fig6}}
\end{figure}

\begin{figure}
\epsscale{1.0}
\plotone{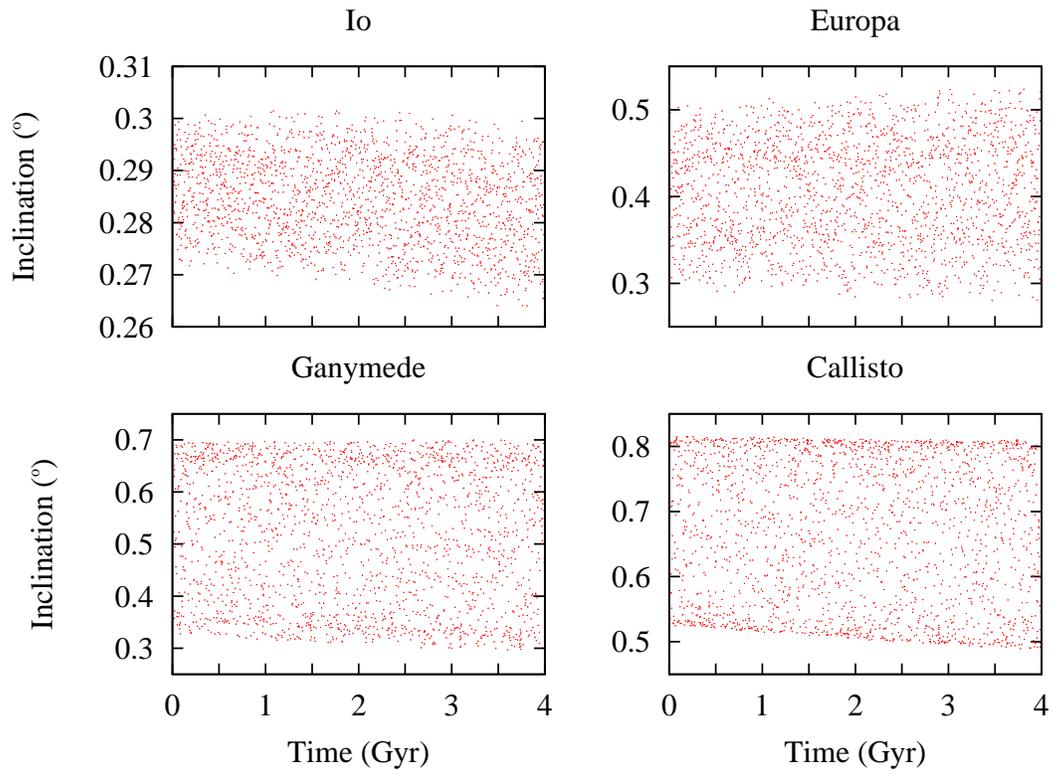}
\caption{Evolution of the orbital inclination (with respect to Jupiter's equatorial plane) of the Galilean 
satellites during the tidal evolution.\label{fig7}}
\end{figure}

Figure \ref{fig6} shows the eccentricity of the four Galilean satellites affected by tides in 4 Gyr.
Most notably, as previously envisioned, we find that Callisto's eccentricity can be significantly damped through the coupling to 
the inner moons resonant system. In the specific case shown in Fig. \ref{fig6}, Callisto's eccentricity
decreases from 0.05 to 0.025. This change should not be taken at its face value because the parameters 
of the tidal model are uncertain. The eccentricity drop can be a factor of several smaller, if the effective 
$k_2/Q$ of Io was lower than considered here, or a factor of several larger, if the effective $k_2/Q$ 
was higher. What this shows is that Callisto's eccentricity {\it could} have been significantly damped
by tides over 4 Gyr. Conversely, Figure \ref{fig7} shows that the orbital inclination of the satellites 
does not appreciably change during the tidal evolution. 

These results have major implications for the interpretation of our scattering experiments discussed in 
the previous sections. First, they show that Callisto's eccentricity cannot be used as a rigid constraint 
on the instability models, because even if Callisto's eccentricity was significantly excited during planetary
encounters, it could have decreased to the present value during the subsequent evolution. The orbital 
inclinations of the moons, on the other hand, represent a more useful constraint on the planetary encounters, 
and can be used to rule out specific instability models. For example, $Case~2$ discussed above is clearly
implausible because Callisto's inclination remains reasonably low only in a few percent of trials.  

As for the semimajor axes of the satellites, after several experiments with their different initial 
values we determined that the global result does not change much by changing semimajor axes. 
So, while the semimajor axes of the Galilean satellites can somewhat change in some cases, 
we believe that it should be enough to point out that tidal migration of Callisto was small 
and insufficient to resolve the problem in Fig. \ref{fig9}.

\subsection{Laplace resonance} \label{laplace}

Another useful constraint on the instability models is represented by the fact that Io, Europa and Callisto 
are in the Laplace resonance. This is because, if the semimajor axes of these satellites change as a result
of the planetary encounters, the orbits can end up escaping from the resonance. If this happens,  
the resonant angle $\phi_L=\lambda_I-3\lambda_E+2\lambda_G$, where $\lambda_I$, $\lambda_E$, and $\lambda_G$, are 
the mean longitudes of Io, Europa, and Ganymede, respectively, starts to circulate. By analyzing our simulations
we found that this happened in 47\%, 93\% and 0\% of the trials in $Cases$ 1, 2 and 3, respectively.

\begin{figure}
\epsscale{1.0}
\plotone{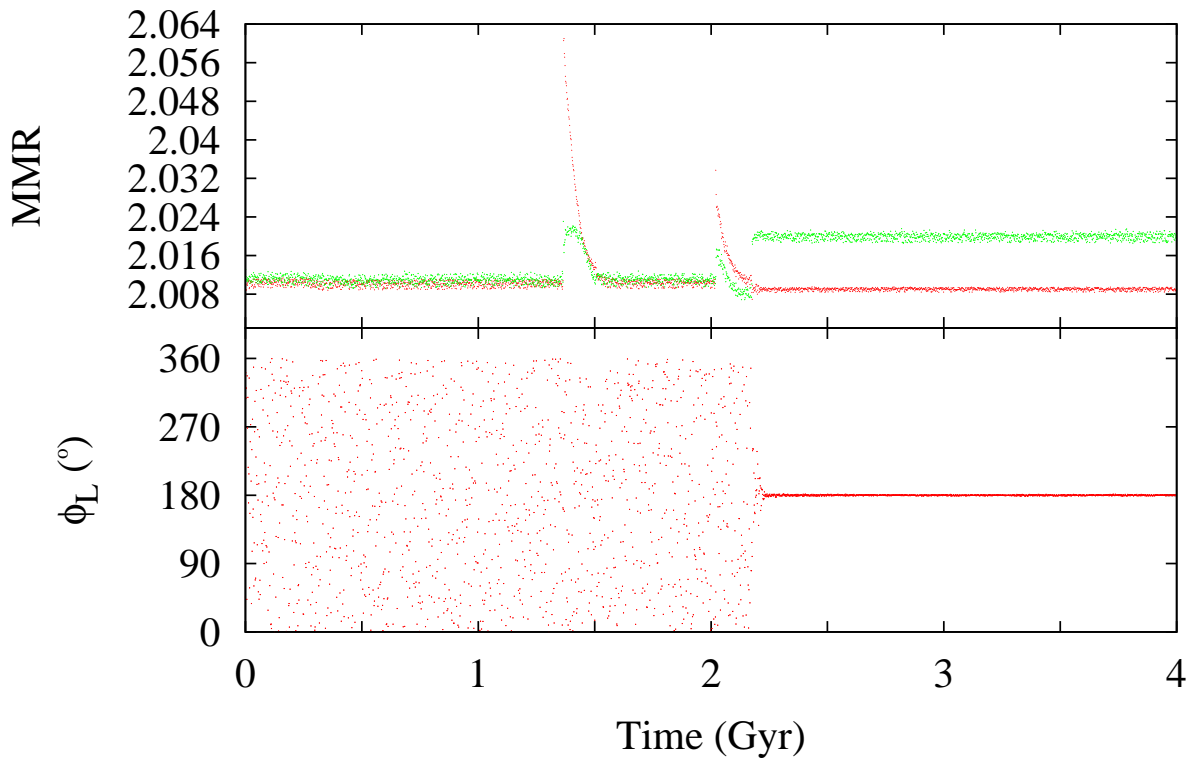}
\caption{Top: Evolution of the mean motion ratios of Io and Europa (red), and Europa and Ganymede (green).
Bottom: Evolution of the Laplace resonance angle $\phi_L=\lambda_I-3\lambda_E+2\lambda_G$. This figure illustrates
that it is possible break the Laplace resonance during planetary encounters, such that $\phi_L$ circulates
at the beginning of our tidal simulations, and re-capture the inner three moons into the Laplace resonance 
when Io moves out by tides. In this specific case, the capture occurred at $\simeq$2.2 Gyr when $\phi_L$ started
librating around $180\degr$. 
\label{fig8}}
\end{figure}

These results show that $Case~2$ is problematic, as already pointed out above based on considerations related 
to the orbital inclination of Callisto. The Laplace resonance constraint, however, is not as solid as the inclination
constraint, because tides could re-capture satellites in the Laplace resonance. For example, if the Europa 
and Ganymede have roughly the same probability to be scattered inward and outward, we find that the outward 
tidal evolution of Io could re-capture Europa and Ganymede in the Laplace resonance in up to 25\% of cases. 
This means, for example, that in up to roughly 65\% of trials in $Case~1$, the Laplace resonance could have 
been left undisturbed or was re-established during the subsequent tidal evolution. An illustration of a case where 
the moons were re-captured in the Laplace resonance is shown in Figure \ref{fig8}.

\section{Conclusions} \label{conclusion}

Here we studied the orbital behavior of the Galilean satellites during the dynamical instability in the outer
Solar System. By carrying out numerical simulations we determined the level of orbital perturbation in three 
instability cases taken from \citet{NM12}. The orbital perturbations occur when Jupiter has a close encounter with 
an ice giant, and the gravity of the ice giant acts on the satellite orbits. We also studied the subsequent 
evolution of the surviving satellites due to the tidal effects.

We found that the orbits of the Galilean satellites can be profoundly affected by the planetary encounters 
between Jupiter and an ice giant, especially in the cases where there the encounter distance $0.02$ AU $< d_{enc} 
< 0.03$ AU. The extremely deep encounters with distance $d_{enc} < 0.02$ AU can be clearly ruled out, because
those would lead to strong orbital excitation, collisions of moons, and their removal. On the other hand, 
encounters with $0.03$ AU $< d_{enc} < 0.05$ AU cause only small variations of the orbital elements, and those
with $d_{enc} >0.05$ AU leave the satellite system essentially undisturbed.

The subsequent tidal evolution of the moons, principally due to the tidal dissipation in the Jupiter-Io pair, 
is capable of damping Callisto's orbital eccentricity. Therefore, even if Callisto's eccentricity becomes 
excited by planetary encounters, it can decrease to values similar to Callisto's present eccentricity 
($\simeq0.03$) later. Similarly, even if the Laplace resonance of the inner three moons can be broken during
planetary instability, it can be reestablished later when Io moves tidally outward, and recaptures Europa
and Ganymede in outer MMRs. Together, this means that the eccentricity and Laplace resonance do not represent
stringent constraints that could be used to rule out specific instability models.

Conversely, the orbital inclinations of the Galilean moons, and mainly that of Callisto, provide an important 
constraint. This is because the inclinations are nearly unaffected by tides. Therefore, if the inclinations
would have been strongly excited during planetary encounters, such as in the $Case~2$ discussed earlier, they
would survive to the present. These cases could clearly be ruled out when compared to the presently low orbital 
inclinations of the Galilean satellites. Specifically, we find that any planetary encounters with $d_{enc} < 0.02$ 
AU ($Case~2$) can clearly be ruled out, and that encounters with $0.02$ AU $< d_{enc} < 0.03$ AU ($Case~1$) 
cannot be too many. 

Encounters with $d_{enc} > 0.03$ AU can happen more often. This is interesting because these more distant encounters
can be important for capture of the irregular satellites at Jupiter \citep{NVD14}. We therefore find that the Galilean 
and irregular satellites represent different constraints on the planetary encounters: the former limit the distance 
and number of a few closest encounters, while the latter require that many distant encounters occur. These constraints
are complementary and should ideally be used together when evaluating the plausibility of a specific instability
model. For example, applying these constraints in Cases 1, 2 and 3 studied here and in \citet{NVD14} we find that
Case 3 is clearly plausible, Case 2 is clearly implausible, and Case 1 stands in the middle, where most parameters
are acceptable except for Callisto's inclination that ends up too high in most trials.  

Still, orbital inclinations could have been damped by another mechanism such as, 
for example, dynamical friction exerted by a disk of debris (Deienno et al. 2012; %\citet{Deienno12}, 
Deienno \& Yokoyama, $in$ $preparation$). 
In this case, Callisto could interact with a disk of debris and its orbital inclination be damped together the other orbital elements. 
Such disk could have formed if one considers that Jupiter could have had more satellites 
in the past than it has today (beyond Callisto's orbit), and those extra satellites were lost by
 collisions among themselves during the instability phase of the Solar System 
(as proposed for Uranus' satellites by \citet{Deienno11}), or by a residual mass produced 
from collisional grinding of irregular satellites \citep{Bottke10}. 
This could be an interesting alternative, because as shown by \citet{Cuk05}, 
dynamical friction is more effective in damping Triton's orbit than tides alone (working on inclination as well). 
However, such a hypothesis needs to be carefully analyzed to determine whether or not the 
mass and life time of the disk are plausibly large to affect Callisto's inclination. 
We leave this work for future investigations.  

In a broader scope, while the Galilean satellite constraints discussed here can be used to rule out specific instability
models, they cannot be used to rule out the jumping-Jupiter model in general. This is because many (but not all) 
previously-developed jumping-Jupiter models should satisfy the Galilean satellite constraint as well. 
There is also the possibility that the initial giant planet system had more than five planets, 
where the additional planets were smaller in mass \citep{NM12}. 
In this case, satellite excitation would be presumably less severe than found in $Cases$ $1$ and $2$ in this paper.
This result should 
be seen in positive light, because the jumping-Jupiter model is required from the terrestrial planet constraint 
\citep{Brasser09,Morbidelli07,NM12}, the dynamical structure of the asteroid belt \citep{Morbidelli10}, and for the capture 
of the irregular satellites at Jupiter \citep{NVD14}. 

% %%%%%%%%%%%%%%%%%%%%%%%%%%%%
\acknowledgments

This work was supported by FAPESP (grants 2012/23732-4 and 2010/11109-5). D.N. was supported by NASA's Outer Planet 
Research program. The work of D.V. was partly supported by the Czech Grant Agency (grant P209-13-013085). 
T. Y. was supported by CNPq. We thank an anonymous referee for useful comments on the submitted manuscript.

% %%%%%%%%%%%%%%%%%%%%%%%%%%%%
%\appendix
%
% %%%%%%%%%%%%%%%%%%%%%%%%%%%%

% %%%%%%%%%%%%%%%%%%%%%%%%%%%%

% %%%%%%%%%%%%%%%%%%%%%%%%%%%%

\clearpage

 %%%%%%%%%%%%%%%%%%%%%%%%%%%%

\clearpage

 %%%%%%%%%%%%%%%%%%%%%%%%%%%%

\clearpage

 %%%%%%%%%%%%%%%%%%%%%%%%%%%%

\clearpage

 %%%%%%%%%%%%%%%%%%%%%%%%%%%%

\clearpage

 %%%%%%%%%%%%%%%%%%%%%%%%%%%%

\clearpage

 %%%%%%%%%%%%%%%%%%%%%%%%%%%%

\clearpage

 %%%%%%%%%%%%%%%%%%%%%%%%%%%%

\clearpage

 %%%%%%%%%%%%%%%%%%%%%%%%%%%%

\clearpage

 %%%%%%%%%%%%%%%%%%%%%%%%%%%%

\clearpage

% %%%%%%%%%%%%%%%%%%%%%%%%%%%%
%Table...
%
%\clearpage
%

% %%%%%%%%%%%%%%%%%%%%%%%%%%%%

\begin{thebibliography}{}

\bibitem[Agnor \& Lin(2012)]{AL12} Agnor, C.~B., \& Lin, D.~N.~C.\ 2012, \apj, 745, 143
\bibitem[Bottke et al.(2010)]{Bottke10} Bottke, W.~F., Nesvorn\'y, D., Vokrouhlick\'y, D., \& Morbidelli, A.\ 2010, \aj, 139, 994
\bibitem[Brasser et al.(2009)]{Brasser09} Brasser, R., Morbidelli, A., Gomes, R., Tsiganis, K., \& Levison, F. H.\ 2009 \aap, 134, 1790
\bibitem[Canup \& Ward(2002)]{Canup02} Canup, R.~M., \& Ward, W.~R.\ 2002, \aj, 124, 3404
\bibitem[Canup \& Ward(2006)]{Canup06} Canup, R. M., \& Ward, W. R.\ 2006, \nat, 441, 834 
\bibitem[Canup \& Ward(2009)]{Canup09} Canup, R.~M., \& Ward, W.~R.\ 2009, University of Arizona Press, 59
\bibitem[Chambers(1999)]{Chambers99} Chambers, J. E.\ 2012, \mnras, 304, 793
\bibitem[\'Cuk \& Gladman(2005)]{Cuk05} \'Cuk, M., \& Gladman, B.~J.\ 2005, \apjl, 626, L113
\bibitem[Deienno et al.(2011)]{Deienno11} Deienno, R., Yokoyama, T., Nogueira, E. C., Callegari, N. Jr., \& Santos, M. T.\ 2011, \aap, 536, A57
\bibitem[Deienno et al.(2012)]{Deienno12} Deienno, R., Yokoyama, T., \& Prado, A.~F.~B.~A.\ 2012, AAS/Division for Planetary Sciences Meeting Abstracts, 44, \#415.09
\bibitem[Fernandez \& Ip(1996)]{FI96} Fernandez, J. A., \& Ip, W. H.,\ 1996, \planss, 44, 431
\bibitem[Gomes et al.(2005)]{Gomes05} Gomes, R. S., Tsiganis, K., Morbidelli, A., \& Levison, H. F.\ 2005, \nat, 435, 466
\bibitem[Hahn \& Malhotra(1999)]{HM99} Hahn, J. M., \& Malhotra, R.\ 1999, \aj, 117, 3041
\bibitem[Lainey et al.(2009)]{Lainey09} Lainey, V., Arlot, J., Karatekin, \"O., \& Hoolst, T. V.\ 2009, \nat, 459, 957
\bibitem[Levison \& Duncan (1994)]{Levison94} Levison, H. F., \& Duncan, M. J.\ 1994, \icarus, 108, 18
\bibitem[Levison et al.(2008)]{Levison08} Levison, H. F., Morbidelli, A., Vanlaerhoven, C., Gomes, R., \& Tsiganis, K.\ 2008, \icarus, 196, 258
\bibitem[Levison et al.(2009)]{Levison09} Levison, H. F., Bottke, W. F., Gounelle, M.,\ et al. 2009, \nat, 460, 364
\bibitem[Levison et al.(2011)]{Levison11} Levison, H.~F., Morbidelli, A., Tsiganis, K., Nesvorn\'y, D., \& Gomes, R.\ 2011, \aj, 142, 152
\bibitem[Lovis et al.(2011)]{Lovis11} Lovis, C., S\'egransan, D., Mayor, M., et al.\ 2011, \aap, 528, A112
\bibitem[Nesvorn\'y et al.(2007)]{Nesvorny07} Nesvorn\'y, D., Vokrouhlick\'y, D., \& Morbidelli, A.\ 2007, \aj, 133, 1962
\bibitem[Nesvorn\'y \& Vokrouhlick\'y(2009)]{NV09} Nesvorn\'y, D., \& Vokrouhlick\'y, D.\ 2009, \aj, 137, 5003
\bibitem[Nesvorn\'y(2011)]{Nesvorny11} Nesvorn\'y, D.\ 2011, \apj, 742, 22
\bibitem[Nesvorn\'y \& Morbidelli(2012)]{NM12} Nesvorn\'y, N. \& Morbidelli, A.\ 2012, \aj, 144, 117
\bibitem[Nesvorn\'y et al.(2013)]{NVM13} Nesvorn{\'y}, D., Vokrouhlick{\'y}, D., \& Morbidelli, A.\ 2013, \apj, 768, 45
\bibitem[Nesvorn{\'y} et al.(2014a)]{NVD14} Nesvorn\'y, D., Vokrouhlick\'y, D., \& Deienno, R.\ 2014a, \apj, 784, 22
\bibitem[Nesvorn\'y et al.(2014b)]{NVD14b} Nesvorn\'y, D., Vokrouhlick\'y, D., Deienno, R., \& Walsh, K. J.\ 2014b, \aj, submitted
\bibitem[Mignard(1979)]{Mignard79} Mignard, F.\ 1979, The Moon and the Planets, 20, 301
\bibitem[Minton \& Malhotra(2011)]{MM11} Minton, D. A., \& Malhotra, R.\ 2011, \apj, 732, 53
\bibitem[Morbidelli et al.(2005)]{Morbidelli05} Morbidelli, A., Levison, H. F., Tsiganis, K., \& Gomes, R.\ 2005, \nat, 435, 462
\bibitem[Morbidelli et al.(2007)]{Morbidelli07} Morbidelli, A., Tsiganis, K., Crida, A., Levison, F. H., \& Gomes, R.\ 2007, \apj, 134, 1790
\bibitem[Morbidelli et al.(2010)]{Morbidelli10} Morbidelli, A., Brasser, R., Gomes, R., Levison, H. F., \& Tsiganis, K.\ 2010, \aj, 140, 1391
\bibitem[Murray \& Dermott(1999)]{Murray99} Murray, C. D., \& Dermott, S. F.\ 1999, Solar System Dynamics, Cambridge University Press, Cambridge
\bibitem[Tremaine et al.(2009)]{tremaine09} Tremaine, S., Touma, J., \& Namouni, F.\ 2009, \aj, 137, 3706
\bibitem[Tsiganis et al.(2005)]{Tsiganis05} Tsiganis, K., Gomes, R. S., Morbidelli, A., \& Levison, H. F.\ 2005, \nat, 435, 459
\bibitem[Van Laerhoven \& Greenberg(2013)]{Van13} Van Laerhoven, C., \& Greenberg, R.\ 2013, \apj, 778, 182
\bibitem[Walsh et al.(2011)]{Walsh11} Walsh, K. J., Morbidelli, A., Raymond, S. N., O'Brien, D. P. \& Mandell, A. M.\ 2011, \nat, 475, 206
\bibitem[Ward \& Canup(2006)]{wardcanup06} Ward, W. R., \& Canup, R. M.\ 2006, \apj, 640, L91
\bibitem[Yoder \& Peale(1981)]{Yoder81} Yoder, C.~F., \& Peale, S.~J.\ 1981, \icarus, 47, 1

\end{thebibliography}
\end{document}